\documentclass[twocolumn,floatfix,preprintnumbers,citeautoscript,newabstract,nofootinbib,a4paper]{revtex4}

\usepackage{amsmath}
\usepackage{amssymb}
\usepackage{graphicx}
\usepackage{bbm}

\newcommand{\be}{\begin{equation}}  
\newcommand{\ee}{\end{equation}}  
\renewcommand{\ol}[1]{\overline{#1}}

\newcommand{\vev}[1]{\langle #1 \rangle}

\newcommand{\SU}[1]{\ensuremath{\mathrm{SU}(#1)}}

\newcommand{\into}{\ensuremath{\;\rightarrow\;}}
\newcommand{\tr}{\operatorname{tr}}

\renewcommand{\tensor}{\ensuremath{\otimes}}

\begin{document}

\preprint{DESY 10-195}
\title{Metastable supersymmetry breaking without scales}

\author{Felix Br\"ummer}
\email{felix.bruemmer@desy.de}

\affiliation{Deutsches Elektronen-Synchrotron DESY\\
Notkestra\ss e 85, D-22607 Hamburg, Germany}

\begin{abstract}
\noindent We construct new examples of models of metastable $D=4$ $N=1$ supersymmetry breaking in which all scales are generated dynamically. Our models rely on Seiberg duality and on the ISS mechanism of supersymmetry breaking in massive SQCD. Some of the electric quark superfields arise as composites of a strongly coupled gauge sector. This allows us to start with a simple cubic superpotential and an asymptotically free gauge group in the ultraviolet, and end up with an infrared effective theory which breaks supersymmetry dynamically in a metastable state. 
\end{abstract}

\maketitle

\section{Introduction}

Low-energy supersymmetry (SUSY) can alleviate the electroweak hierarchy problem in two distinct senses. First, softly broken SUSY protects the Higgs potential of the supersymmetric Standard Model from quadratic divergences. The hierarchy between the electroweak scale and some high fundamental scale, e.g.~the grand-unified scale or the Planck scale, is thus stabilized against radiative corrections. Second, if SUSY is unbroken at tree-level, then by the renormalization theorem it can at most be broken by non-perturbative effects. These are typically exponentially small in units of the fundamental scale, so one may hope to explain not just the stability, but also the origin of the hierarchy without appealing to the details of the UV physics.\footnote{One might of course disregard the second point altogether and be content with a model of tree-level SUSY breaking whose dimensionful parameters happen to be many orders of magnitude below fundamental scale, relegating an explanation to the unknown UV completion. In this Letter we shall however take seriously the prospect of understanding the origin of the hierarchy purely within effective field theory.}

This second point serves as a major motivation to study models of dynamical SUSY breaking (DSB) \cite{Witten:1981nf, Affleck:1983mk}. In DSB models, the SUSY breaking hidden sector contains a gauge theory which becomes strongly coupled at some infrared scale $\Lambda$. While the tree-level superpotential preserves SUSY, non-perturbative effects such as instantons or gaugino condensation can generate additional terms leading to SUSY breakdown. The SUSY breaking scale will then involve $\Lambda$, which can naturally be many orders of magnitude below the fundamental scale. 

The models we will be analysing in this Note are DSB models relying on the now classic ISS mechanism of SUSY breaking \cite{Intriligator:2006dd}. An essential ingredient of the ISS mechanism is Seiberg duality \cite{Seiberg:1994bz}: Certain asymptotically free gauge theories, most notably supersymmetric QCD with suitable matter content, are dual to infrared-free gauge theories at energies below their strong-coupling scale $\Lambda$. Under this duality a superpotential mass term $\mu\,q\tilde q$ for the elementary matter superfields $q$, $\tilde q$ turns into a linear term $\mu\Lambda\,M$ for the composite infrared field $M$. The presence of this linear term eventually triggers SUSY breaking in a metastable vacuum at $M=0$.

In a strict sense the ISS mechanism does not offer a fully dynamical explanation of why the scale of SUSY breaking is small, since $\mu\ll\Lambda$ must be put in by hand (for $\mu\gtrsim\Lambda$ the matter fields would decouple before the theory can become strongly coupled). Several models have been constructed to remedy this situation, generating $\mu$ from strong gauge dynamics of some auxiliary sector \cite{Dine:2006gm,Aharony:2006my,Brummer:2007ns}. Here we take the idea somewhat further by constructing models whose $q$ degrees of freedom are themselves composites of a strongly coupled gauge sector. We are thus making use of two strong-coupling transitions, by which an originally cubic superpotential term in the UV is mapped first to a quadratic and then to a linear operator in the effective IR theory.

Schematically, we will proceed as follows. Consider the gauge group $\SU{n}\times\SU{N}$ and matter superfields $\tilde q$, $\Phi$, and $Q$ transforming as $\ol\Box\tensor{\bf 1}$, $\Box\tensor\ol\Box$, and ${\bf 1}\tensor\Box$ respectively. This allows for a marginal operator
\be\label{sketchyW}
W=\lambda\, Q\Phi\tilde q\,,
\ee 
where $\lambda$ is a dimensionless coupling. The $\SU{N}$ factor becomes strongly coupled at a scale $\Lambda_N$. In certain cases its infrared dynamics can be described by means of a different, weakly coupled dual theory involving a composite ``meson'' field $q\sim Q\Phi/\Lambda_N$. Then the effective superpotential will contain a term
\be
W_{\rm eff}=\lambda\Lambda_N\,q\tilde q+\ldots
\ee 
which corresponds to a mass $\mu=\lambda\Lambda_N$ for $q$ and $\tilde q$. At a scale $\Lambda_n$ the $\SU{n}$ factor becomes strongly coupled, and in the far infrared we end up with the desired linear term for the composite $M\sim q\tilde q/\Lambda_n$,
\be
W_{\rm eff}=\lambda\Lambda_N\Lambda_n\, M+\ldots
\ee
The coefficient $\lambda\Lambda_N\Lambda_n$ eventually sets the scale of SUSY breaking; it does not involve any fundamental mass parameters, so supersymmetry is broken truly dynamically. The marginal parameter $\lambda$ must however be chosen small enough to guarantee $\lambda\Lambda_N<\Lambda_n$.

It is the aim of this Letter to flesh out the above construction in detail. In its simplest version (which is the one we are concerned with), with a single bifundamental $\Phi$, there are strong constraints on the remaining matter content and on the ranks of the gauge groups. The only choices which do not suffer from instabilities turn out to require $N=F=f$, where $F$ and $f$ are the overall number of $\SU N$ and $\SU n$ flavours respectively, and $N>n$ (with $N=n$ also potentially allowed but uncalculable).

\section{General framework}\label{prel}

The basic ingredients of the models we are investigating are the gauge group $\SU{n}\times\SU{N}$, where we take $n\leq N$ without loss of generality, and some matter fields allowing for a superpotential resembling Eq.~\eqref{sketchyW}. We thus introduce a bifundamental field $\Phi$ along with $f$ copies of $\SU{n}$ antiquarks $\tilde q$ and $F$ copies of $\SU{N}$ quarks $Q$. We also allow for $\SU n$ fundamentals $p$ and $\SU N$ antifundamentals $\widetilde P$. 
The matter content is summarized in Table \ref{generaltable}.
\begin{table}[t]
\begin{center}
\begin{tabular}{c||c|c||c|c|c|c}
& $\SU{N}$ & $\SU{n}$ & $\SU{F}$ & $\SU{F-n}$ & $\SU{f}$ & $\SU{f-N}$ \\ \hline
$\Phi$ & $\ol\Box$ & $\Box$ & $\mathbf{1}$ & $\mathbf{1}$ & $\mathbf{1}$ & $\mathbf{1}$ \\
$Q$ & $\Box$ & $\mathbf{1}$ & $\ol\Box$ & $\mathbf{1}$ & $\mathbf{1}$ & $\mathbf{1}$ \\
$\tilde q$ & $\mathbf{1}$ & $\ol\Box$ & $\mathbf{1}$ & $\mathbf{1}$ & $\Box$ & $\mathbf{1}$  \\
$\widetilde P$ & $\ol\Box$ & $\mathbf{1}$ & $\mathbf{1}$ & $\Box$ & $\mathbf{1}$ & $\mathbf{1}$ \\
$p$ & $\mathbf{1}$ & $\Box$ & $\mathbf{1}$ &  $\mathbf{1}$ & $\mathbf{1}$ & $\ol\Box$ \\
\end{tabular}
\caption{\label{generaltable} Gauge symmetries and non-abelian flavour symmetries for vanishing superpotential. The symmetries in the four rightmost columns are global, while $\SU N$ and $\SU n$ are gauged.}
\end{center}
\end{table}
It is restricted by the absence of gauge anomalies: Anomaly cancellation for $\SU{n}$ requires $f\geq N$, with $f-N$ spectator fields $p$ present if $f>N$. Analogous statements hold for $\SU N$. One could include more $\SU N\times\SU n$ bifundamentals, or fields in larger representations, but we will refrain from that for now. 

A superpotential will explicitly break some of the flavour symmetries of Table \ref{generaltable}. The most general renormalisable superpotential respecting the gauge symmetries is
\be\label{electricWplus}
W=\lambda^I_j\,Q_I\Phi\tilde q^j+\mu^I_A\, Q_I\widetilde P^A+m^b_j\,p_b\tilde q^j
\ee
where $\lambda^I_j$ is an $F\times f$ matrix, $\mu^I_A$ is an $F\times(F-n)$ matrix, $m^b_j$ is an $f\times(f-N)$ matrix, and where we have suppressed gauge indices.\footnote{For $N=2,3$ or $n=2,3$ one could also include renormalisable baryonic operators, without affecting our conclusions.} In keeping with the principle that all dimensionful parameters should be of the order of the fundamental scale, the last two terms in Eq.~\eqref{electricWplus} will only lead to the decoupling of ${\rm rank}\,(\mu^I_A)$ pairs of $Q,\widetilde P$ and ${\rm rank}\,(m^b_j)$ pairs of $p,\tilde q$ in the UV. We will therefore omit them (and redefine the fields and parameters accordingly). Tree-level masses for the remaining fields can be forbidden by a $\mathbb{Z}_3$ or $R$-symmetry, or by imposing that $\SU{F-n}$ and $\SU{f-N}$ should be classically preserved. The superpotential we are working with is then
\be\label{electricW}
W=\lambda^I_j\,Q_I\Phi\tilde q^j\,.
\ee

Flavour rotations permit a singular value decomposition of $\lambda^I_j$, bringing it into the form
\be\begin{split}\label{lambdas}
(\lambda^I_j)=\left(\begin{array}{ccc}\lambda_1 & &  \\ & \ddots &  \\ & & \lambda_f \\ \hline & &\\ & 0 & \\ & &  \end{array} \right)\;
\text{or}\;(\lambda^I_j)=\left(\begin{array}{ccc|ccc}\lambda_1 & & & & &\\ & \ddots & & & \text{\raisebox{3pt}{0}} &\\ & & \lambda_F &&&\end{array} \right)
\end{split}\ee
(depending on whether $f<F$ or $f\geq F$). For simplicity we will from now on consider the case where all $\lambda_i$ are equal, $\lambda_1=\lambda_2=\ldots\equiv\lambda$; it is straightforward to extend the analysis to the more general case where all $\lambda_i$ are nonzero and of similar magnitude. Situations where there are large hierarchies among the $\lambda_i$, or where some of the $\lambda_i$ vanish, are less interesting as will become clear below.

Both gauge factors are required to be asymptotically free, so $3N>F$ and $3n>f$. They run to strong coupling in the infrared. We now distinguish two cases, namely $\SU N$ becoming strongly coupled at higher energies than $\SU n$ and vice versa.

\subsection{The case $\Lambda_N>\Lambda_n$}

Suppose first that $\SU N$ becomes strongly coupled at a scale $\Lambda_N$, where $\SU n$ gauge couplings are still negligible. We are then dealing with $\SU N$ SQCD with $F$ flavours of quarks $Q$ and antiquarks $\Phi,\,\widetilde P$. SQCD with $N$ colours and $F$ flavours has a known and calculable weakly coupled infrared description if $N\leq F<\frac{3}{2}\,N$ in terms of its Seiberg dual theory \cite{Seiberg:1994bz}, so we restrict ourselves to the case where $N$ and $F$ are in this range. Some of the mesonic degrees of freedom $q$ in the dual theory can be identified with the $Q\Phi$ composites, $q_I\sim Q_I\Phi/\Lambda_N$. Here we have absorbed a factor $\Lambda_N$ in the definition of the fields $q$ so that they have canonical dimension. Below the scale $\Lambda_N$ the degrees of freedom are those of the magnetic dual of $\SU N$, and the $\SU N$ singlet fields we started with. In the infrared, where the magnetic gauge dynamics becomes negligible, they constitute an $\SU n$ gauge theory with $F+f-N$ flavours. The effective superpotential contains a term
\be\label{Wmag}
W_{\rm eff}=\lambda^I_j\Lambda_N\,q_I\tilde q^j+\ldots
\ee 
descending from Eq.~\eqref{electricW}. It gives a supersymmetric mass $\lambda\Lambda_N$ to rank$\,(\lambda^I_j)$ pairs of quarks and antiquarks.

The infrared structure of this theory does not appear to be very interesting at first sight, since for $\lambda$ of order one the massive $\SU n$ flavours should be integrated out, and one is left with an $\SU n$ gauge theory with massless matter. Instead we now keep all quarks light by dialling $\lambda\ll 1$. The $\SU n$ gauge coupling will become strong at some lower scale $\Lambda_n$; if $\lambda$ is chosen such that $\lambda\Lambda_N<\Lambda_n$, all flavours will remain dynamical to below that scale. Supposing that $\SU n$ is in the free magnetic range as well, we end up with the ISS model of metastable SUSY breaking. After another Seiberg duality on the $\SU n$ factor we obtain a low-energy effective theory which breaks SUSY in a metastable vacuum.

There are two possible issues here. First, choosing $\lambda$ small may be regarded as problematic in view of naturalness. We stress however that $\lambda$ is merely a marginal parameter, as opposed to a relevant one. Furthermore, the tuning is used to overcome the discrepancy between the two dynamically generated scales $\Lambda_n$ and $\Lambda_N$, which could in principle be close together. The naturalness problem we eventually aim to solve with dynamical supersymmetry breaking, by comparison, involves the potentially much larger hierarchy between the SUSY breaking scale and the UV completion scale.

Second, for generic choices of $F$, $f$, $N$ and $n$ one ends up with both massive and massless $\SU n$ flavours. This is because the superpotential Eq.~\eqref{Wmag} gives masses to only $F$ or $f$ of the $F+f-N$ quarks and antiquarks (and to even less if some of the $\lambda_i$ in Eq.~\eqref{lambdas} are zero). An effective ISS model with both massive and massless flavours still gives rise to supersymmetry breaking at tree-level. However, some of the pseudo-moduli will no longer be stabilized at one-loop \cite{Franco:2006es}. Indeed it was found in \cite{Giveon:2008wp} that along the directions where this happens, the two-loop contribution to the effective potential causes a runaway towards the supersymmetric vacuum. A runaway may also appear if there are large hierarchies between the quark masses, in which case the tachyonic two-loop contributions from the heavier quarks may overwhelm the one-loop contributions from the lighter ones. It has subsequently been argued \cite{Giveon:2008ne} that higher-dimensional operators in the superpotential may stabilize these runaways away from the point of maximal unbroken symmetry, and that phenomenologically promising metastable vacua may appear as a consequence (for some recent developments see e.g.~\cite{SchaferNameki:2010mg}). While such ideas may lead to interesting generalizations when applied to our model, for now our aim is to build a hidden sector that is by itself as UV-complete as possible. We should therefore demand that the quark mass matrix is square and has full rank, such that all quarks become massive.

This, together with the restrictions $F\geq n$, $f\geq N$ (from anomaly cancellation) and $N\geq n$ (by convention) translates into the condition
\be
F=f=N.
\ee 
The resulting models are discussed in Sects.~\ref{working} and \ref{spec}.

\subsection{The case $\Lambda_n>\Lambda_N$}
The discussion is in large parts analogous to the preceding one. For $n\leq f<\frac{3}{2}\,n$ we can dualize the $\SU n$ gauge factor at the scale $\Lambda_n$. Below we obtain an effective $\SU N$ gauge theory with $F+f-n$ flavours. The superpotential Eq.~\eqref{electricW} gives rise to a mass term. Requiring that all flavours acquire a mass, and that this mass is below $\Lambda_N$, implies that $\lambda$ has to be chosen sufficiently small and that
\be
F=f=N=n.
\ee 
We thus obtain an even stricter condition than before. The resulting model is discussed in Sect.~\ref{spec}.

\section{Models in the stable range}\label{working}
Postponing the case $n=N$ until later, we now investigate models with $\Lambda_N>\Lambda_n$ and $F=f=N>n$ in detail. More precisely, to have $\SU n$ in the free magnetic range we demand $n< N<\frac{3}{2}\,n$. The matter content and non-abelian symmetries are summarised in Table \ref{FfNtable}.
\begin{table}
\begin{center}
\begin{tabular}{c||c|c||c|c}
& $\SU{n}$ & $\SU{N}$ & $\SU{N}$ & $\SU{N-n}$ \\ \hline
$\Phi$ & $\Box$ & $\ol\Box$ & $\mathbf{1}$ & $\mathbf{1}$  \\
$\tilde q$ & $\ol\Box$ & $\mathbf{1}$ & $\Box$ & $\mathbf{1}$ \\
$Q$ & $\mathbf{1}$ & $\Box$ & $\ol\Box$ & $\mathbf{1}$  \\
$\widetilde P$ & $\mathbf{1}$ & $\ol\Box$ & $\mathbf{1}$ &  $\Box$ \\
\end{tabular}
\end{center}
\caption{\label{FfNtable}Matter content and symmetries for $F=f=N>n$. The first two symmetries $\SU n$ and $\SU N$ are gauge symmetries. The $\SU N$ and $\SU{N-n}$ in the last two columns represent classically unbroken global symmetries. The flavour $\SU N$, in particular, corresponds to the diagonal subgroup of $\SU F\times\SU f$ in Table \ref{generaltable}.}
\end{table}
The unique renormalisable superpotential is
\be
W=\lambda\,Q\Phi\tilde q.
\ee

To the extent that the $\SU{n}$ gauge dynamics can be neglected at the scale $\Lambda_N$, the model is just $\SU{N}$ SQCD with $N$ flavours of quarks $Q$ and antiquarks $\Phi,\widetilde P$. This theory has a low-energy description where the quarks confine into a meson $M=q\oplus\tilde p$, a baryon $B$, and an antibaryon $\widetilde B$. Here $q$ and $\tilde p$ correspond to the composites $q=Q\Phi/\Lambda_N$ and $\tilde p=Q\widetilde P/\Lambda_N$. The remaining symmetries act as in Table \ref{magtable}.
\begin{table}
\begin{center}
\begin{tabular}{c||c||c|c}
& $\SU{n}$ & $\SU{N}$ & $\SU{N-n}$ \\ \hline
$q$ & $\Box$  & $\ol\Box$ & $\mathbf{1}$ \\
$\tilde q$ & $\ol\Box$ & $\Box$ & $\mathbf{1}$  \\
$\widetilde p$ & $\mathbf{1}$ &  $\ol\Box$ &  $\Box$ \\
$B$ & $\mathbf{1}$ &  $\mathbf{1}$ & $\mathbf{1}$ \\
$\widetilde B$ & $\mathbf{1}$ & $\mathbf{1}$ &  $\mathbf{1}$ \\
\end{tabular}
\end{center}
\caption{\label{magtable}Matter content and symmetries below the $\SU N$ confinement scale. $\SU n$ is gauged, while $\SU N$ and $\SU{N-n}$ are global.}
\end{table}

The fields are subject to the quantum-deformed moduli space constraint  
\be\label{defmodcons}
\frac{\det M}{(\Lambda_N)^N}-\frac{B\widetilde B}{(\Lambda_N)^2}=1.
\ee
The effective superpotential reads
\be\label{Wcons}
W_{\rm eff}=\lambda\Lambda_N\,q\tilde q+(\Lambda_N)^2 T\left(\frac{\det M}{(\Lambda_N)^N}-\frac{B\widetilde B}{(\Lambda_N)^2}-1\right).
\ee
$T$ is a Lagrange multiplier to enforce the constraint Eq.~\eqref{defmodcons}. Note that here and in the following we are omitting uncalculable prefactors of order one. We will consider low-energy fluctuations on the baryonic branch of this theory, where the constraint is satisfied with $B\widetilde B=-(\Lambda_N)^2$ and $\det M=0$. The K\"ahler potential for such fluctuations is approximately canonical \cite{Intriligator:1996pu}. 

For $\lambda$ of order one, $q$ and $\tilde q$ will decouple around the scale $\Lambda_N$ and the gauge degrees of freedom will form a pure super-Yang-Mills theory which does not break SUSY. We instead choose $\lambda$ sufficiently small, such that $\lambda\Lambda_N\ll\Lambda_n$, so $q$ and $\tilde q$ remain as light flavours of $\SU{n}$. The $\SU n$ gauge coupling will now become strong at the lower scale $\Lambda_n$. Since we chose $n$ and $N$ in the range $n< N<\frac{3}{2}\,n$, this theory has an infrared-free magnetic dual in terms of a $\SU{N-n}$ gauge theory with $N$ flavours of dual quarks $\chi$ and antiquarks $\tilde\chi$ and a meson $\widetilde M$. The $\SU{n}$ singlet degrees of freedom remain as spectators. The field content in the infrared is given in Table \ref{magmagtable},
\begin{table}
\begin{center}
\begin{tabular}{c||c||c|c}
& $\SU{N-n}$ & $\SU{N}$ & $\SU{N-n}$ \\ \hline
$\widetilde M$ & $\mathbf{1}$ & $\ol\Box\tensor\Box$ & $\mathbf{1}$ \\
$\tilde p$ & $\mathbf{1}$& $\ol\Box$ & $\Box$ \\
$B$ & $\mathbf{1}$ & $\mathbf{1}$ & $\mathbf{1}$\\
$\widetilde B$& $\mathbf{1}$ & $\mathbf{1}$ & $\mathbf{1}$\\
$\chi$ & $\Box$ & $\ol\Box$ & $\mathbf{1}$ \\
$\tilde\chi$ & $\ol\Box$ & $\Box$ & $\mathbf{1}$ \\
\end{tabular}
\end{center}
\caption{\label{magmagtable}Field content after a Seiberg duality transformation on $\SU n$. Only the $\SU{N-n}$ listed on the left is gauged; it is the dual magnetic gauge group of $\SU n$.}
\end{table}
and the superpotential becomes
\be\begin{split}\label{Wcons2}
&W_{\rm eff}=\lambda\Lambda_N\Lambda_n\,\tr\widetilde M+\chi\widetilde M\tilde\chi\\&+\Lambda_N^2\,T\left(\frac{\sqrt{N!}}{(N-n)!}\frac{(\Lambda_n)^{2n-N}}{(\Lambda_N)^N}\,\det(\tilde\chi\tilde p)-\frac{B\widetilde B}{(\Lambda_N)^2}-1\right).
\end{split}
\ee
Here we have neglected an Affleck-Dine-Seiberg term which is irrelevant for small meson expectation values. We have also expressed $\det M$ from Eq.~\eqref{Wcons} in terms of the magnetic degrees of freedom: Under Seiberg duality, baryonic operators constructed from $q$ are mapped to baryonic operators constructed from $\tilde\chi$ in the magnetic theory. In particular,
\be\begin{split}
\frac{1}{n!}\epsilon^{i_1\ldots i_n\,j_1\ldots j_{N-n}}\epsilon_{c_1\ldots c_n}\frac{
q^{c_1}_{i_1}\cdots q^{c_n}_{i_n}}{(\Lambda_n)^n}\;\leftrightarrow\\
\frac{\sqrt{N!}}{(N-n)!}\epsilon^{\alpha_1\ldots\alpha_{N-n}} \frac{\tilde\chi_{\alpha_1}^{j_1}\cdots\tilde\chi_{\alpha_{N-n}}^{j_{N-n}}}{(\Lambda_n)^{N-n}}
\end{split}
\ee
(where the $i_k$ and $j_l$ are $\SU N$ flavour indices, the $c_i$ are $\SU{n}$ colour indices, and the $\alpha_j$ are $\SU{N-n}$ dual colour indices). Therefore 
\be
\det M=\frac{1}{n!}\epsilon^{i_1\ldots i_n\,j_1\ldots j_{N-n}}\epsilon_{c_1\ldots c_n}
q^{c_1}_{i_1}\cdots q^{c_n}_{i_n} \tilde p^{1}_{j_1}\cdots \tilde p^{{N-n}}_{j_{N-n}}
\ee 
is mapped into
\be\begin{split}
&\frac{\sqrt{N!}}{(N-n)!}\frac{1}{(\Lambda_n)^{N-2n}}\epsilon^{\alpha_1\ldots\alpha_{N-n}} (\tilde p_{j_1}^1\tilde \chi^{j_1}_{\alpha_1})\cdots (\tilde p_{j_{N-n}}^{N-n}\tilde\chi^{j_{N-n}}_{\alpha_{N-n}})\\
&\qquad=\frac{\sqrt{N!}}{(N-n)!}\frac{1}{(\Lambda_n)^{N-2n}}\det(\tilde p\tilde\chi)\,.
\end{split}
\ee

On the baryonic branch the constraint Eq.~\eqref{defmodcons} can be satisfied by setting $B\widetilde B=-(\Lambda_N)^2$ and $\tilde p=0$. Then $W_{\rm eff}$ in Eq.~\eqref{Wcons2} becomes precisely the magnetic superpotential of the ISS model, which is well-known to break supersymmetry in a metastable state. The vacuum energy is
\be
\vev V=\,n\,|\lambda\Lambda_n\Lambda_N|^2.
\ee
All scales are generated by dimensional transmutation. The only small parameter, $\lambda$, is dimensionless. 

The need for a small marginal parameter is common in models which use the ISS mechanism in SUSY-breaking models without scales \cite{Brummer:2007ns, Essig:2007xk}. To gain some intuition on a realistic upper bound on $\lambda$, let us construct a limiting case for which our approximations can still be considered reliable. Take for instance $n=8$, $N=10$, and assume that the $\SU n$ coupling is $g_n\approx 1$ at the scale $\Lambda_N$; with this value $\SU n$ is still perturbative but on the brink of strong coupling. With $g_n(\Lambda_N)=1.1$ and the naive one-loop estimate
\be
g_n^2(\Lambda_n)=\frac{g^2_n(\Lambda_N)}{1-\frac{3n-N}{8\pi^2}g_n^2(\Lambda_N)\log(\Lambda_N/\Lambda_n)}\,,
\ee
the scale $\Lambda_n$ where $g_n$ diverges is about two orders of magnitude below $\Lambda_N$. For $\Lambda_N=10^8$ GeV and $\Lambda_n=10^6$ GeV, with $\lambda=10^{-3}$, one obtains a SUSY breaking scale of a few $\cdot\;10^5$ GeV, in the correct range for low-scale gauge mediation. Of course, there are always uncalculable ${\cal O}(1)$ factors involved (which may or may not work in our favour). To really trust our model it seems more reasonable to demand at least $\lambda\lesssim 10^{-4}$, also since the vacuum becomes more long-lived for small $\lambda$. 

The model still contains various massless fields, which should be decoupled if it is to serve as a realistic hidden sector. To this end we can introduce an additional gauge singlet field $S$, transforming as $\Box\otimes\ol\Box$ under the $\SU N\times\SU{N-n}$ flavour symmetry. This allows for an operator $Q\widetilde P S$ in the UV superpotential, so that after the $\SU N$ strong-coupling transition $\tilde p$ and $S$ obtain a mass $\sim\Lambda_N$. As an additional benefit the model is now forced to be on the baryonic branch. A flat direction remains in the baryon sector, corresponding to a rescaling $B\into e^b B$ and $\widetilde B\into e^{-b}\widetilde B$. We need to assume that uncalculable K\"ahler terms will stabilize this direction sufficiently close to the symmetric point $B=\widetilde B=i\Lambda_N$ (parametrically large values for either $|B|$ or $|\widetilde B|$ are unacceptable, since they would lead to a loss of control over higher-dimensional operators coupling baryons to other fields).

We are now in a position to couple our model to the visible sector. There are many models of direct gauge-mediated or messenger gauge-mediated SUSY breaking, relying on ISS-like metastable vacua (for some examples, see \cite{massg}). It would be especially interesting to see if the common problems of direct gauge mediation (such as the gaugino mass problem and the Landau pole problem) can somehow be overcome in an extension of our model, possibly even while preserving its nice UV properties. For now we leave this issue for future work.  

It is instructive to see how the model fails to give rise to a metastable vacuum in the case $\Lambda_n>\Lambda_N$. Below the scale $\Lambda_n$ we should use the Seiberg dual of $\SU{n}$, which is a $\SU{N-n}$ magnetic gauge theory. It contains a meson $\widetilde Q=\Phi\tilde q/\Lambda_n$ and two dual quarks which we call again $\chi$ and $\tilde\chi$. The full field content is listed in Table \ref{ngtrNtable1}.
\begin{table}
\begin{center}
\begin{tabular}{c||c|c||c|c}
& $\SU{N-n}$ & $\SU N$ & $\SU{N}$ & $\SU{N-n}$ \\ \hline
$\widetilde Q$ & $\mathbf{1}$ & $\ol\Box$ & $\Box$ & $\mathbf{1}$  \\
$Q$ & $\mathbf{1}$ & $\Box$ & $\ol\Box$ & $\mathbf{1}$  \\
$\chi$ & $\Box$ & $\mathbf{1}$ & $\ol\Box$ &  $\mathbf{1}$ \\
$\tilde\chi$ & $\ol\Box$ & $\Box$ & $\mathbf{1}$ &  $\mathbf{1}$ \\
$\widetilde P$ & $\mathbf{1}$ & $\ol\Box$ & $\mathbf{1}$ &  $\Box$ \\
\end{tabular}
\end{center}
\caption{\label{ngtrNtable1}The field content for $\Lambda_n>\Lambda_N$, after a Seiberg duality transformation on $\SU n$. As before, the first two columns represent gauge symmetries and the last two columns represent global flavour symmetries.}
\end{table} 
The superpotential becomes, up to Affleck-Dine-Seiberg terms generated by the dual gauge dynamics,
\be
W_{\rm eff}=\chi\widetilde Q\tilde\chi+\lambda\Lambda_n\,Q\widetilde Q.
\ee

If $\lambda$ is of order one, then $Q$ and $\widetilde Q$ decouple supersymmetrically at the scale $\Lambda_n$. The remaining fields form an $\SU{N-n}\times\SU{N}$ gauge theory with massless matter and no superpotential, which does not break SUSY.

Considering instead again the case that $\lambda\ll 1$, so that $Q$ and $\widetilde Q$ are kept light up to scales $<\Lambda_N$, the theory becomes effectively SQCD with $N$ colours and $2N-n$ flavours of quarks $Q$, $\tilde\chi$ and antiquarks $\widetilde Q$, $\widetilde P$. It is in the free magnetic range, which is easily checked to follow from $n<N<\frac{3}{2}\,n$. It therefore has a Seiberg dual description at energies below $\Lambda_N$ in terms of mesons \mbox{$M=Q\widetilde Q/\Lambda_N$}, \mbox{$\tilde\eta = \tilde\chi\widetilde Q/\Lambda_N$}, $\eta=Q\widetilde P/\Lambda_N$, and \mbox{$\zeta=\tilde\chi\widetilde P/\Lambda_N$}, as well as magnetic quarks $\rho,\sigma$ and antiquarks $\tilde\rho,\tilde\sigma$. There is a \mbox{$\SU{(2N-n)-N}=\SU{N-n}$} magnetic gauge symmetry, in addition to the $\SU{N-n}$ gauge symmetry which is the magnetic gauge symmetry of the first duality transformation, and a \mbox{$\SU{N}\times\SU{N-n}$} flavour symmetry. The symmetry properties of the various fields are listed in Table \ref{ngtrNtable2}.
\begin{table}
\begin{center}
\begin{tabular}{c||c|c||c|c|c|c}
& $\SU{N-n}$ & $\SU{N-n}$ & $\SU{N}$ & $\SU{N-n}$   \\ \hline
$M$ & $\mathbf{1}$ & $\mathbf{1}$ & $\ol\Box\otimes\Box$ & $\mathbf{1}$\\
$\tilde\eta$ & $\ol\Box$ & $\mathbf{1}$ & $\Box$ & $\mathbf{1}$ \\
$\eta$ & $\mathbf{1}$ & $\mathbf{1}$ & $\ol\Box$ & $\Box$ \\
$\zeta$ & $\ol\Box$ & $\mathbf{1}$ & $\mathbf{1}$ & $\Box$  \\
$\chi$ & $\Box$ & $\mathbf{1}$ & $\ol\Box$ & $\mathbf{1}$  \\
$\rho$ & $\mathbf{1}$ & $\Box$& $\ol\Box$ & $\mathbf{1}$  \\
$\tilde\rho$ & $\mathbf{1}$ & $\ol\Box$& $\Box$ &  $\mathbf{1}$  \\
$\sigma$ & $\mathbf{1}$ & $\Box$ & $\mathbf{1}$ & $\ol\Box$  \\
$\tilde\sigma$ & $\Box$ & $\ol\Box$ &$\mathbf{1}$ &$\mathbf{1}$  \\
\end{tabular}
\end{center}
\caption{\label{ngtrNtable2}Fields and symmetries after a second Seiberg duality transformation, now on $\SU N$. Only the first two $\SU{N-n}$ factors are gauged (but weakly coupled in the IR).}
\end{table} 
The superpotential is
\be
\begin{split}
W_{\rm eff}=&\lambda\Lambda_N\Lambda_n\,\tr M+\Lambda_N\,\chi\tilde\eta\\
&+\rho M\tilde\rho+\rho\tilde\eta\tilde\sigma+\sigma\eta\tilde\rho+\sigma\zeta\tilde\sigma\,.
\end{split}
\ee
In the far infrared, $\chi$ and $\tilde\eta$ decouple. Supersymmetry is broken at tree-level near the origin of field space by the rank condition, by the $F$-terms of $M$: The rank of $\rho\tilde\rho$ is $N-n$, while the rank of $\partial(\tr M)/\partial M_i^j$ is $N$. The $\SU{N}$ electric quark mass matrix (or equivalently the linear term in the meson) does not have full rank in flavour space, however, since there is no linear term for $\zeta$. As mentioned in Sect.~\ref{prel}, this will destabilize the SUSY-breaking point at two-loop level \cite{Giveon:2008wp}.

\section{The uncalculable case $F=f=N=n$}\label{spec}
\begin{figure} 
\begin{center}
\includegraphics[width=20mm]{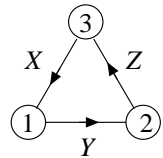}
\end{center}
\caption{\label{triangle} The $F=f=N=n$ quiver. Circles with label $i$ represent $\SU{N}_i$ symmetries, and arrows represent bifundamental chiral superfields. Node $3$ may or may not be gauged.}
\end{figure}

We next turn to the case $F=f=N=n$, adapting our notation slightly for convenience. The non-abelian symmetry group is $\SU{N}_1\times\SU{N}_2\times\SU{N}_3$. The third factor arises as the diagonal subgroup of the $\SU F\times\SU f$ in Table \ref{generaltable} which is preserved by $W$. Since it is free of anomalies, it could be gauged; however its gauge dynamics will play no role in the following discussion. The matter fields are now called $X$, $Y$, and $Z$. The model is summarized by the quiver diagram in Fig.~\ref{triangle}.
We label the $\SU N$ factors according to their coupling strengths in the UV. The superpotential is
\be
W=\lambda\,XYZ.
\ee
We choose $\lambda$ such that the strong-coupling scales satisfy $\Lambda_1>\Lambda_2>\lambda\Lambda_1>\Lambda_3$. Then, at the highest scale $\Lambda_1$, $\SU{N}_1$ confines and $X$ and $Y$ combine into a meson $M=XY/\Lambda_1$. The superpotential becomes
\be\label{Wsu23}
W_{\rm eff}=\lambda\Lambda_1\,MZ+\Lambda_1^2\,T\left(\frac{\det M}{\Lambda_1^N}-\frac{B\widetilde B}{\Lambda_1^2}-1\right).
\ee
This is massive SQCD, where $M$ and $Z$ are $N$ flavours of the $\SU{N}_2$ gauge factor (again provided that $\SU{N}_3$ is negligibly weakly coupled at scales above $\Lambda_2$). The quark mass term is small because $\lambda$ is chosen small. The extra baryon degrees of freedom satisfy a deformed moduli space constraint, which should however be unimportant for the IR dynamics of the meson near the baryonic branch. At the scale $\Lambda_2$, $\SU{N}_2$ confines, and $M$ and $Z$ combine into a meson $\widetilde M=MZ/\Lambda_2$. The superpotential becomes
\be\begin{split}
W_{\rm eff}=&\lambda\Lambda_1\Lambda_2\tr \widetilde M+\Lambda_1^2\,T\left(\frac{\Lambda_2^{N-1}\;b}{\Lambda_1^N}-\frac{B\widetilde B}{\Lambda_1^2}-1\right)\\
&+\Lambda_2^2\,\widetilde T\left(\frac{\det\widetilde M}{\Lambda_2^N}-\frac{b\tilde b}{\Lambda_2^2}-1\right).
\end{split}
\ee
The theory has reduced to a ``Polonyi model'' for $\tr\widetilde M$, with additional constraint terms. Whether or not there is a metastable vacuum near $\widetilde M=0$ depends on uncalculable higher-dimensional operators in the effective K\"ahler potential. The status of this model is therefore similar to the ISS model for equal numbers of flavours and colours, where the existence of a metastable vacuum has been conjectured \cite{Intriligator:2006dd}, but never definitely established.

Note that we have identified the $\SU{N}_2$ baryon $b$ with the $\SU{N}_1$ meson determinant. This implies that, if the theory is on the baryonic branch of $\SU{N}_2$, it will be slightly displaced from the baryonic branch of $\SU{N}_1$.

In any case, the true vacua of the theory are supersymmetric and located at
\be
b=\tilde b=0,\qquad\widetilde M=\Lambda_2\mathbbm{1},\qquad B\widetilde B=-\Lambda_1^2.
\ee
Even if the conjectured metastable vacuum exists, it should somehow be prevented from decaying into the true vacuum too quickly for the model to be viable.

It is amusing to note that one may recover the model of Sect.~\ref{working} from the model of this Section, by breaking $\SU{N}_2$ explicitly to $\SU{N-n}\times\SU{n}$ and taking the limit of negligible $\SU{N-n}$ gauge coupling. The correspondence between the matter fields is then $X\simeq Q$, $Y\simeq\Phi\oplus\widetilde P$, and $Z\simeq\tilde q\oplus S$.

\section{Conclusions}

In summary, we have built a model of dynamical metastable SUSY breaking without a fundamental scale, by taking $\SU n$ SQCD whose quarks are the composites of an $\SU{N}$ gauge group with $n<N<\frac{3}{2}\,n$. The strong-coupling scale $\Lambda_N$ of $\SU{N}$ should be higher than the strong-coupling scale $\Lambda_n$ of the original SQCD gauge group $\SU{n}$, and the overall number of $\SU N$ flavours should be $N$. The tree-level superpotential contains a single cubic term, whose coefficient $\lambda$ has to be tuned small to keep the composite quarks light at scales below both strong-coupling transitions. Supersymmetry is broken in a metastable state by the ISS mechanism. The choice $N=n$ might also give metastable SUSY breaking, depending on uncalculable terms in the K\"ahler potential. In future work, it should be interesting to study how our model can be extended to couple to the visible sector, and if it can be used to construct a realistic model of gauge-mediated supersymmetry breaking.

\bigskip

\section*{Acknowledgements}

I would like to thank Mark Goodsell for useful discussions, and Joerg Jaeckel for helpful comments on the manuscript.

\end{document}